# Brownian Motion in a Speckle Light Field: Tunable Anomalous Diffusion and Deterministic Optical Manipulation

Giorgio Volpe[a,1], Giovanni Volpe[b] & Sylvain Gigan[a]

a. Institut Langevin, UMR7587 of CNRS and ESPCI ParisTech, 1 rue Jussieu, 75005 Paris, France

b. Physics Department, Bilkent University, Cankaya, 06800 Ankara, Turkey

1. Corresponding author: giorgio.volpe@espci.fr

## Abstract

The motion of particles in random potentials occurs in several natural phenomena ranging from the mobility of organelles within a biological cell to the diffusion of stars within a galaxy. A Brownian particle moving in the random optical potential associated to a *speckle*, i.e., a complex interference pattern generated by the scattering of coherent light by a random medium, provides an ideal mesoscopic model system to study such phenomena. Here, we derive a theory for the motion of a Brownian particle in a speckle and, in particular, we identify its universal characteristic timescale levering on the universal properties of speckles. This theoretical insight permits us to identify several interesting unexplored phenomena and applications. As an example of the former, we show the possibility of tuning anomalous diffusion continuously from subdiffusion to superdiffusion. As an example of the latter, we show the possibility of harnessing the speckle *memory effect* to perform some basic *deterministic* optical manipulation tasks such as guiding and sorting by employing *random* speckles, which might broaden the perspectives of optical manipulation for real-life applications by providing a simple and cost-effective technique.

**Introduction**

Various phenomena rely on particles performing stochastic motion in random potentials. Examples range from the nanoscopic world of molecules undergoing anomalous diffusion within the cytoplasm of a cell [1] to the Brownian motion of stars within galaxies [2]. Another example of this kind of phenomena is given by the motion of a Brownian particle in a random optical potential generated by a speckle, i.e., the random light field resulting from complex light scattering in optically complex media, such as biological tissues, turbid liquids and rough surfaces (see background in Figure 1a-c) [3-4]. This latter example is particularly suited to work as a model system because its parameters (e.g, particle size and material, illumination light) are easily controllable and its dynamics are easily accessible by standard optical microscopy techniques [5]. Earlier experimental works showed the possibility of trapping particles in high-intensity speckle light fields [6-9], the simplest optical manipulation task, and the emergence of superdiffusion in an active media constituted by a dense solution of microparticles that generates a time-varying speckle [10]. However, apart from these previous studies, there is little understanding of the interaction of Brownian motion with random light potentials and the intrinsic randomness of speckles is largely considered a nuisance to be minimized for most purposes, e.g., in optical manipulation [11-12].

In this Letter, we study the Brownian motion of a mesoscopic particle in a speckle and, in particular, we derive the characteristic timescale $\tau$ of such motion, which is universal as it depends only on the universal properties of speckle light fields [3, 13-14]. This theoretical insight permits us to identify several potential phenomena and applications of interest, which have not been explored so far. First of all, while the emergence of superdiffusion in time-varying speckles has already been observed [10], we demonstrate that speckles provide a tunable and controllable model system to study anomalous diffusion by relating τ to the emergence of subdiffusion and superdiffusion in time-varying speckles. Moreover, we demonstrate the possibility of harnessing the *memory effect* of speckle fields [13-14] to perform basic deterministic optical manipulation tasks such as guiding and sorting, which go beyond selective optical trapping in high-intensity speckles [7-8].

**Results and discussion**

Speckle light patterns can be generated by different processes, such as scattering of a laser on a rough surface, multiple scattering in an optically complex medium, or mode-mixing in a multimode fiber [4]. In general, they are the result of the interference of a large number of waves propagating along different directions and with a random phase distribution, and, despite their random appearance, they share some universal statistical properties (see Supplementary Note S1) [3, 15]. In particular, the speckle has a negative exponential intensity distribution, and the normalized spatial autocorrelation function $C_I(\Delta \mathbf{r})$ can be approximated by a Gaussian [16]:

$$C_I(\Delta \mathbf{r}) = \frac{\langle I(\mathbf{r}+\Delta \mathbf{r})I(\mathbf{r})\rangle}{\langle I(\mathbf{r})^2\rangle} \approx e^{-\frac{|\Delta \mathbf{r}|^2}{2\sigma^2}}, \qquad (1)$$

where $I(\mathbf{r})$ is the speckle intensity as a function of the position $\mathbf{r}$ and the standard deviation $\sigma$ is proportional to the average speckle grain size.

The motion of a Brownian particle in a static speckle field is the result of random thermal forces and deterministic optical forces. In the following discussion, we focus on Rayleigh particles (see Methods), i.e., particles whose dimensions are smaller than the light wavelength λ, but our conclusions hold also for larger particles [17]. Optical gradient forces are the dominant deterministic forces acting on Rayleigh particles, and they attract high-refractive index particles towards the intensity maxima of the optical field [17]. Optical scattering forces push the particles in the direction of propagation of light, so that in the presence of a boundary, such as the glass surface of a microchannel, they effectively confine the particles in a quasi two dimensional space.

As a particle moves in a speckle, the optical force acting on it changes both in magnitude and direction with a characteristic time scale $\tau = \frac{L}{\langle v \rangle}$, where $L$ is the correlation length of the optical force field and $\langle v \rangle$ is the average particle drift speed. The optical force field correlation function (Figure 1d) is

$$C_F(\Delta \mathbf{r}) = \frac{k^2}{\sigma^2} \langle I \rangle^2 \left(2 - \frac{|\Delta \mathbf{r}|^2}{\sigma^2}\right) e^{-\frac{|\Delta \mathbf{r}|^2}{2\sigma^2}}, \qquad (2)$$

so that $L = \sqrt{2}\sigma$ (see detailed derivation in Methods and Supplementary Note S2). Since the particle motion is overdamped [18], the average particle drift speed is $\langle v \rangle = \frac{\langle F \rangle}{\gamma}$, where $\gamma$ is the particle friction coefficient and $\langle F \rangle \approx \frac{k}{\sigma} \langle I \rangle$ is the average force (see Supplementary Note S2). Thus, we obtain

$$\tau \approx \sqrt{2} \frac{\sigma^2 \gamma}{k \langle I \rangle}, \qquad (3)$$

which is our central theoretical result and permits us to identify the range where various interesting phenomena and applications take place. In the following, after considering how $\tau$ is related to the Brownian motion of a particle in a static speckle, we will present two such examples via numerical experiments.

We start by considering the motion of a particle in a static speckle. As shown by the trajectory in Figure 1a, when the optical forces are relatively low (average force $\langle F \rangle = 10$ fN), the particle is virtually freely diffusing. As the forces increase (Figure 1b, $\langle F \rangle = 50$ fN), first a subdiffusive behavior emerges where the particle is metastably trapped in the speckle grains, while it can still move between them [19]. Finally, for even stronger forces (Figure 1c, $\langle F \rangle = 200$ fN), the particle remains trapped in one of the speckle grains for a very long time, as previously observed experimentally [6-9]. These observations can be interpreted in terms of $\tau$: for relatively high forces ($\langle F \rangle = 200$ fN, $\tau \approx 5.6$ ms), $\tau$ is quite low which means that the particle, on average, experiences a restoring force towards its previous position quite soon in its motion, having little possibility to escape a speckle grain; for relatively low forces ($\langle F \rangle = 10$ fN, $\tau \approx 112.5$ ms) instead, $\tau$ is much higher which means that the particle has time to diffuse away from a speckle grain before actually experiencing the influence of the optical forces exerted by it. These qualitative considerations can be made more precise by calculating the mean square displacement $\text{MSD}(\Delta t)$ of the particle motion (see Methods). As shown in Figure 1e, for low optical forces and high $\tau$ the mean square displacement is substantially linear in $\Delta t$, i.e. $\text{MSD}(\Delta t) \approx 4D_0\Delta t$, where $D_{\text{SE}}$ is the Stokes-Einstein diffusion coefficient. As the forces increase and $\tau$ decreases, there is a transition towards a subdiffusive regime characterized by $\text{MSD}(\Delta t) \propto \Delta t^\beta$ with $\beta < 1$. We remark that, for very large $\Delta t$, the motion returns diffusive, i.e. $\beta = 1$, albeit with an effective diffusion coefficient $D_{\text{eff}} < D_{\text{SE}}$.

We move now to an example of fundamental interest: the demonstration that using speckles it is possible to control and tune anomalous diffusion continuously from subdiffusion to superdiffusion by employing a *time-varying* speckle which changes over a timescale $\xi$ similar to $\tau$. A time-varying speckle can result from a time-varying environment [10], but can also be produced in a more controllable way by modulating spatially or spectrally the laser that generates it [4]. As can be seen in Figure 2a, different diffusive regimes emerge depending on the value of the ratio $\xi/\tau$, thus allowing one to tune the diffusive behavior of the particle just relying on the external control of the speckle time scale $\xi$. For $\xi/\tau \gg 1$, the speckle motion is adiabatic so that the particle can reach its equilibrium distribution in the optical potential before the speckle changes. This leads to a subdiffusive behavior, i.e., $\text{MSD}(\Delta t) \propto \Delta t^\beta$ with $\beta < 1$, as in a static speckle. For $\xi/\tau \ll 1$, the particle cannot follow the fast variation of the speckle so that the average optical force on the particle is zero leading to a diffusive behavior, i.e., $\text{MSD}(\Delta t) \approx 4D_0 \Delta t$. For $\xi/\tau \approx 1$, the particle is subject to time-varying forces, which can induce a superdiffusive behavior, i.e., $\text{MSD}(\Delta t) \propto \Delta t^\beta$ with $\beta > 1$. As in the case of the static speckle field, at long timescales, the particle motion will become again diffusive. Figure 2b highlights the transition from subdiffusion to superdiffusion by plotting $\frac{D_{\text{eff}}}{D_{SE}}$ as a function of $\xi/\tau$ for various $\langle F \rangle$: if the resulting $D_{\text{eff}} > D_{SE}$, the particle has undergone superdiffusion. In this way, the diffusive behavior of a Brownian particle in a speckle field can be deterministically controlled by tuning the relevant adimensional parameter $\xi/\tau$, thus providing a simple model system to study anomalous diffusion [20], which has been shown to occur naturally, e.g., in the kinetics of single molecules in living cells [1].

The second example is more applied and is the demonstration that it is possible to deterministically control the motion of a Brownian particle by using speckle fields, which sets the stage to perform optical manipulation tasks such as guiding particles in a particular direction, despite the randomness of the illumination. The relevant parameter for guiding is the ratio between the speckle speed $V_s$ and the average drift velocity of the particle in a static speckle $\langle v \rangle = \frac{L}{\tau}$. As $V_s/\langle v \rangle$ increases, the average guiding velocity $\langle v_p \rangle$ reaches a maximum for $V_s/\langle v \rangle \approx 1$, as shown in Figure 3a. A small speckle translation up to a few micrometers can be implemented capitalizing on the speckle property known as memory effect [13, 14]: for a speckle generated by a thin sample, a small tilt of the illumination, easily achievable, e.g., with a galvanometric mirror or an acousto-optic deflector, entails a small spatial translation of the speckle. As shown in Figure 3b, this is sufficient to realize a Brownian ratchet [21]: the speckle repeatedly shifts first slowly ($V_s/\langle v \rangle = 1$) by 1 µm in the positive direction, which exerts a strong drag on the particle, and then fast ($V_s/\langle v \rangle = 10$) back to the initial position, which has little effect on the particle position. In 250 ms, the particle is dragged by $\approx$ 3µm in the direction of the speckle shift, while the particle's trajectory in the perpendicular direction remains unaffected (Figure 3c).

This guiding capability of a speckle pattern can be combined with standard microfluidic systems in order to perform tasks, such as sieving and sorting. For example, a static speckle can be employed to realize a *speckle sieve* (Figure 4a). As a liquid containing 200 nm and 250 nm particles flows from left to right at 42µm/s, a static speckle efficiently holds the larger particles back while the smaller ones go through almost unaffected; interestingly, the size of the particle that are held back can be dynamically adjusted by changing the intensity of the speckle. A spatiotemporal varying speckle instead can be employed to realize a *speckle sorter* (Figure 4b). In a configuration similar to the one for the speckle sieve, the memory effect can be used to exert a perpendicular force to the flow selectively on the larger particles, so that each kind of particle gets guided into a different channel. Using similar configurations, which are routinely used in microfluiduics, it is possible to separate particles on the basis of various parameters, such as their size or their refractive index, as shown in Supplementary Figures S4 and S5; the resolution of this optical fractionation [22-24] is only limited by the size of the speckle, i.e., the longer the speckle the higher the sensitivity in particle's size or refractive index. These devices can, therefore, be scaled to achieve the high throughput or sensitivity needed in microfluidics (thousands of particles per second) by increasing the flow speed and laser power, as it is also the case for alternative optofluidics devices [22-24]. Moreover, an additional advantage of speckles is that they are also intrinsically widefield: a speckle covering a width of a few hundreds of micrometers could sort many particles in parallel in a broader microfluidc chamber, where flow speed is strongly reduced. In fact, speckle light fields with such universal statistical properties are straightforwardly generated over large areas using different processes, such as scattering of a laser on a rough surface, multiple scattering in an optically complex medium, or mode-mixing in a multimode fiber [4].

In conclusion, we have developed a theoretical framework that allows one to convert the randomness of a speckle light field into a tunable tool to deterministically influence the motion of a Brownian particle. In particular, we performed numerical experiments to show the applicability of this concept to the tunable control of anomalous diffusion and to perform standard optical manipulation tasks, such as sorting and guiding. While sorting can be achieved using optical traps or lattices [22-24], the use of the universal properties of speckle fields has the advantage of requiring very simple optical setups and a very low degree of control over the experimental environment, thus being readily compatible with lab-on-a-chip or in-vivo applications inside scattering tissues, through which light propagation naturally leads to speckles.

# Methods

**Simulation of Brownian motion in a speckle pattern.** The motion of a Brownian particle of radius $R$ in a generic force field can be modeled with the following Langevin equation [25]:

$$m\ddot{\mathbf{r}} + \gamma\dot{\mathbf{r}} = \sqrt{2k_BT\gamma}\mathbf{W} + \mathbf{F}(\mathbf{r}), \tag{4}$$

where $\mathbf{r}$, $m$ and $\gamma = 6\pi\eta R$ are respectively the particle's velocity, mass and friction coefficient, $\eta$ the viscosity of the surrounding medium, $\mathbf{W}$ a white noise vector, $k_B$ the Boltzmann constant and $T$ the temperature of the system. For a Rayleigh particle, $\mathbf{F}(\mathbf{r}) = k\boldsymbol{\nabla}I(\mathbf{r})$, where $k = \frac{1}{4}\text{Re}(\alpha)$ and $\alpha$ is the particle polarizability, which depends on the particle's volume, shape and composition [26]. For a spherical particle with radius $R$ and refractive index $n_p$ immersed in a liquid with refractive index $n_m$, $\alpha = 4\pi n_m^2\varepsilon_0 R^3 \frac{n_p^2 - n_m^2}{n_p^2 + n_m^2}$. Inserting the full expression of the force and neglecting inertial effects in the motion of the particle [18], Equation (4) simplifies as:

$$\dot{\mathbf{r}} = \sqrt{2D_{SE}}\mathbf{W} + \frac{k}{\gamma}\boldsymbol{\nabla}I, \tag{5}$$

where $D_{SE} = \frac{k_BT}{\gamma}$ is the Stokes-Einstein's diffusion coefficient of the Brownian particle. Simulations of Brownian motion in the field of forces generated by the speckle were therefore obtained by numerically solving Equation (5) [27]. In all simulations, the particles are polystyrene beads ($n_p = 1.59$) in water ($n_m = 1.33$, $\eta = 0.001$ Ns/m², $T = 300$ K). Due to the low scattering cross-section of the particles and their low concentration, we consider that the force field of the speckle is not influenced by the particle, unless one considers relatively dense samples [10]. Even though we used a 2-dimensional model, our results can be readily extended to a 3-dimensional situation. In the presence of flow, we assumed laminar flow because of the low Reynolds numbers associated to microfluidc channels [18, 28].

**Mean square displacement calculation.** The calculation of the MSD is performed according to $\text{MSD}(\Delta t) = \langle \mathbf{r}^2(\Delta t) \rangle = \int \mathbf{r}^2 P(\mathbf{r}, \Delta t) d^2 \mathbf{r}$, where $P(\mathbf{r}, \Delta t)$ is the probability density function of finding a particle at position $\mathbf{r}$ at time $\Delta t$ [1]. In practice, each MSD curve was obtained by averaging over 500 different particle trajectories that were simulated over 100 s after waiting enough time for the particles to thermalize in the random optical potential given by the speckle.

## Acknowledgements

The authors thank Andrew Griffiths and Patrick Tabeling for insightful discussions. Giovanni Volpe was partially supported by Marie Curie Career Integration Grant (MC-CIG) PCIG11 GA-2012-321726. Sylvain Gigan acknowledges funding from Agence Nationale de La Recherce (ANR-JCJC-ROCOCO), the City of Paris (Programme Emergence) and the European Research Council (under grant N°278025).

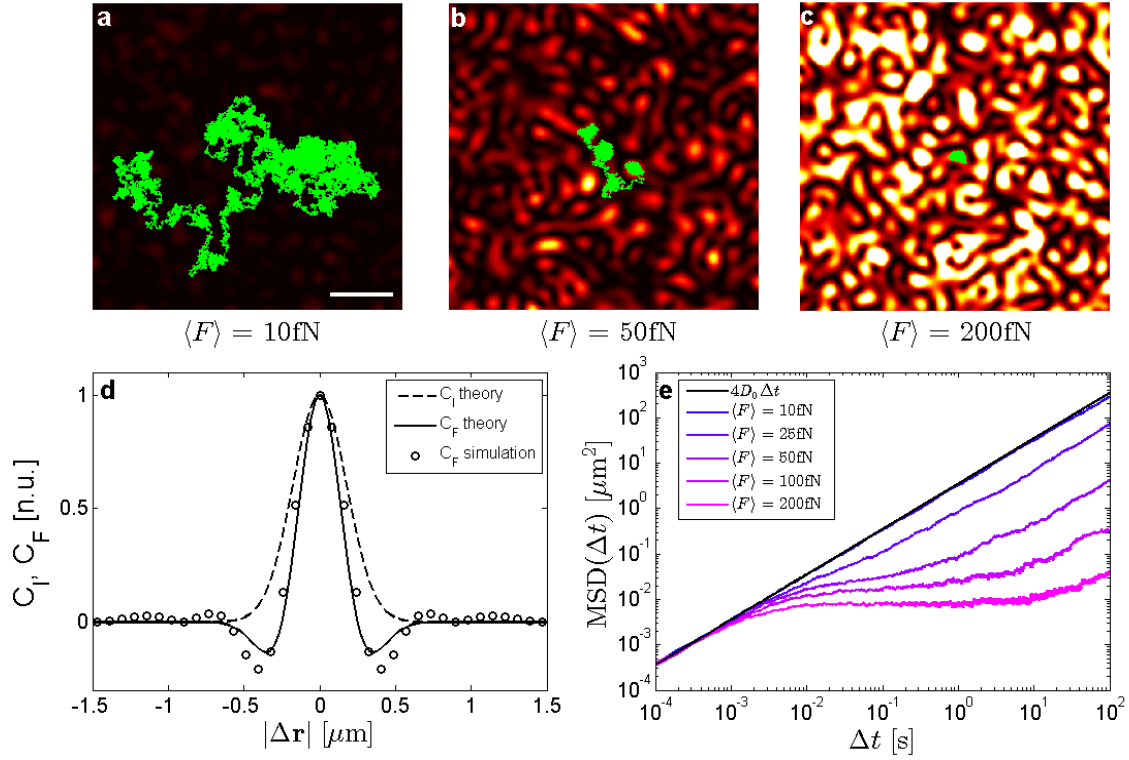

**Figure 1: Subdiffusion in a static speckle pattern.** (**a-c**) The background represents a speckle pattern generated by a circular aperture (λ = 1064 nm, speckle grain 490 nm); the white scale bar corresponds to 2 μm. The trajectories (green solid lines) show progressive confinement of a polystyrene bead (R = 250 nm, $n_p$ = 1.59) in water ($n_m$ = 1.33, $\eta$ = 0.001 Ns/m$^2$, T = 300 K) as a function of the increasing speckle intensity corresponding to an average force on the particle of (**a**) ⟨F⟩ = 10 fN (⟨I⟩ = 13 mW/μm$^2$), (**b**) ⟨F⟩ = 50 fN (⟨I⟩ = 65 mW/μm$^2$), (**c**) ⟨F⟩ = 200 fN (⟨I⟩ = 260 mW/μm$^2$). (**d**) Normalized autocorrelation function of the force field produced by the speckle according to our theoretical model (solid line) and in the simulated speckle pattern (circles). The dashed line represents the theoretical normalized autocorrelation function of the speckle intensity. (**e**) Brownian particle mean square displacements as a function of ⟨F⟩ (purple lines), and their deviation from Einstein's free diffusion law (black line).

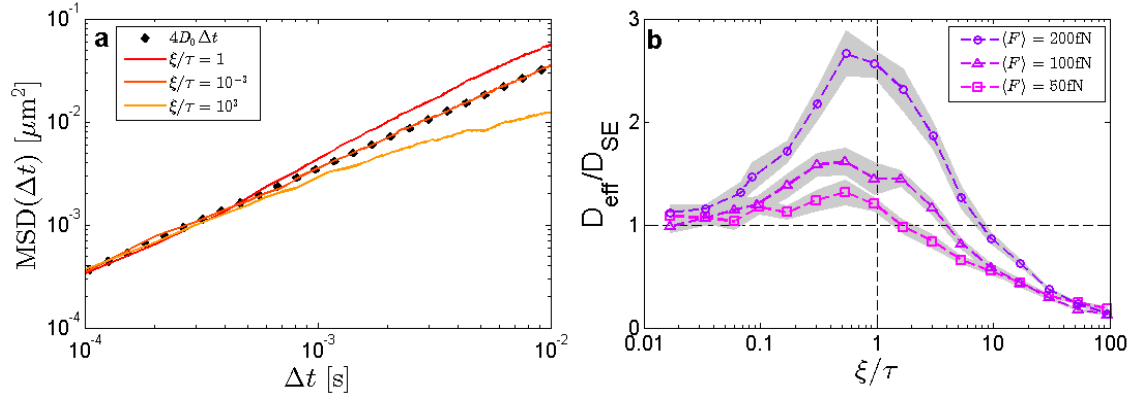

**Figure 2: Superdiffusion in a time varying speckle pattern.** (**a**) Mean square displacements in logarithmic scale for a Brownian particle moving in a speckle pattern which varies on a timescale $\xi \approx \tau$ (red line), $\xi \ll \tau$ (orange line), and $\xi \gg \tau$ (yellow line). The dots represent Einstein's free diffusion law. (**b**) The effective diffusion of the motion at long timescales as a function of $\xi/\tau$ shows a transition from subdiffusion ($D_{\text{eff}} < D_{SE}$) to superdiffusion ($D_{\text{eff}} > D_{SE}$). The maximum value of the superdiffusion appears for $\xi \approx \tau$ ($\tau \approx 22.5$ ms for $\langle F \rangle = 50$ fN, $\tau \approx 11.2$ ms for $\langle F \rangle = 100$ fN and $\tau \approx 5.6$ ms for $\langle F \rangle = 200$ fN). Every mean point is averaged over 500 particle trajectories 100 s long, and whose initial position was randomly chosen within the speckle field. The gray shaded areas represent one standard deviation around the average values.

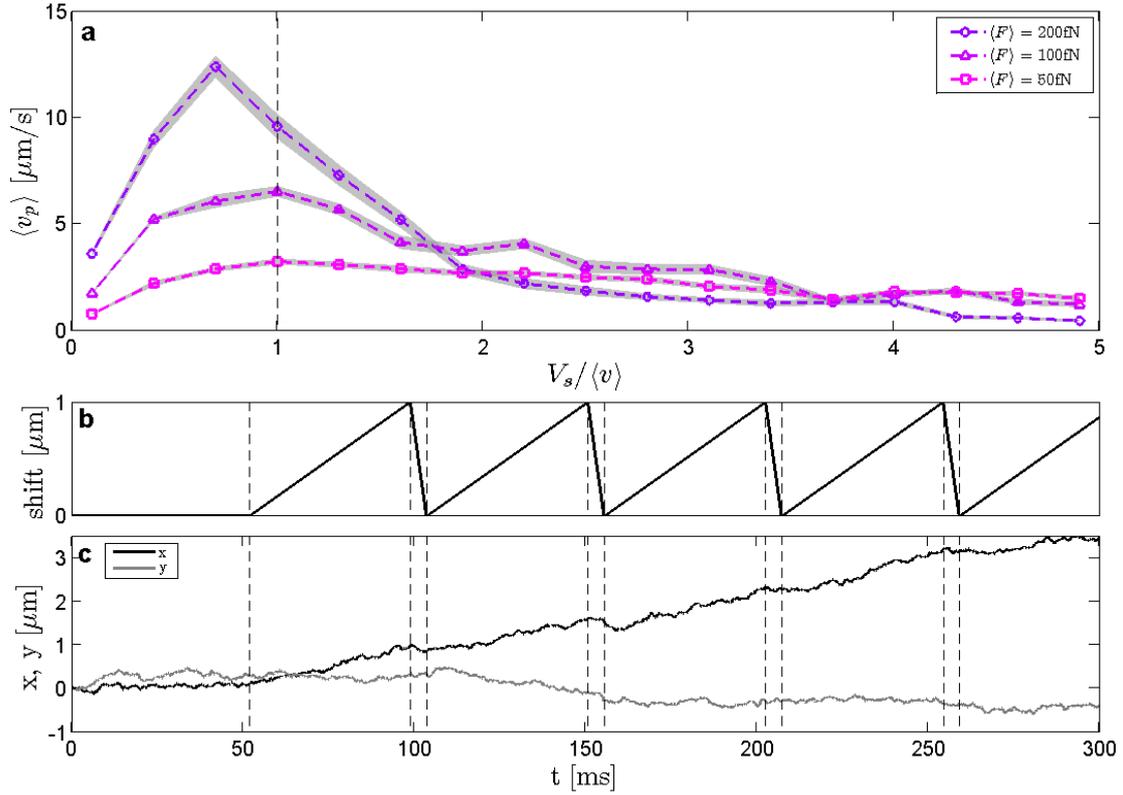

**Figure 3: Guiding by the speckle memory effect.** (**a**) Average guiding velocity $\langle v_p \rangle$ in the direction of the speckle shift as a function of the shift speed $V_s$ for $\langle F \rangle = 200$ fN (circles), $\langle F \rangle = 100$ fN (triangles), and $\langle F \rangle = 50$ fN (squares). The maximum $\langle v_p \rangle$ is achieved for $V_s \approx \langle v \rangle$ ($\langle v \rangle \approx 10.5$ μm/s for $\langle F \rangle = 50$ fN, $\langle v \rangle \approx 21.2$ μm/s for $\langle F \rangle = 100$ fN and $\langle v \rangle \approx 42.4$ μm/s for $\langle F \rangle = 200$ fN). Every mean velocity point is calculated over 500 trajectories simulated during 10 s and whose initial position was randomly chosen within the speckle field. The gray shaded areas represent one standard deviation around the average values. (**b**) Speckle shift and (**c**) particle displacement as a function of time in the direction of the speckle shift $x$ (black line) and in the orthogonal direction $y$ (gray line). The speckle repeatedly shifts first slowly in the positive direction by 1 μm, and then fast to the initial position. The dashed lines represent the instant of time when there is a change of trend in the speckle shift.
16

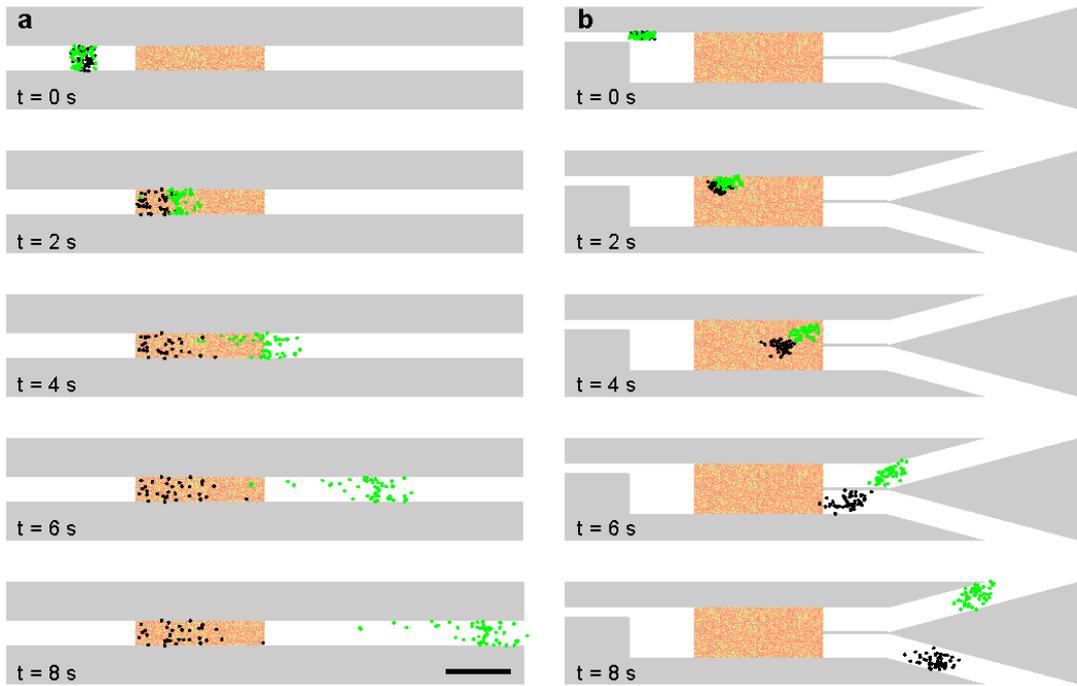

**Figure 4: Microfluidic speckle sieve and speckle sorter.** (**a**) Lapse-time snapshots of the motion of polystyrene particles with radius R = 200 nm ($\langle F \rangle$ = 90 fN, green dots) and R = 250 nm ($\langle F \rangle$ = 46 fN, black dots) in a microfluidic speckle sieve, where a static speckle (red shaded area) traps the smaller particles while it lets the larger particles go away with the flow (flow speed 42 μm/s). (**b**) Same as in (**a**), but with the speckle ratcheting in the direction orthogonal to the flow by 1 μm (flow speed 34 μm/s). The black scale bar represents 50 μm.

# Supplementary Information

# Brownian Motion in a Speckle Light Field: Tunable Anomalous Diffusion and Deterministic Optical Manipulation

Giorgio Volpe[a,1], Giovanni Volpe[b] & Sylvain Gigan[a]

*a. Institut Langevin, UMR7587 of CNRS and ESPCI ParisTech, 1 rue Jussieu, 75005 Paris, France*

*b. Physics Department, Bilkent University, Cankaya, 06800 Ankara, Turkey*

*1. Corresponding author: giorgio.volpe@espci.fr*

**Contents:**

**Supplementary Note S1 – Speckle generation and properties**

**Supplementary Note S2 – Optical forces in a speckle field**

**Supplementary Figure S1 – Speckle intensity distribution**

**Supplementary Figure S2 – Speckle autocorrelation function**

**Supplementary Figure S3 – Optical force distribution**

**Supplementary Figure S4 – Fractionation by size**

**Supplementary Figure S5 – Fractionation by refractive index**

## Supplementary Note S1 – Speckle generation and properties

*Speckle intensity distribution*

A speckle is an interference figure resulting from random scattering of coherent light by a complex medium. The probability density function of the speckle intensity $I$ follows the negative exponential distribution [3, 15]:

$$p(I) = \frac{1}{\langle I \rangle} e^{-\frac{I}{\langle I \rangle}}, \quad (S1)$$

where $\langle I \rangle$ is the average speckle intensity. Figure S1 shows the very good agreement between theoretical and numerical distributions of the speckle intensities used in the simulations.

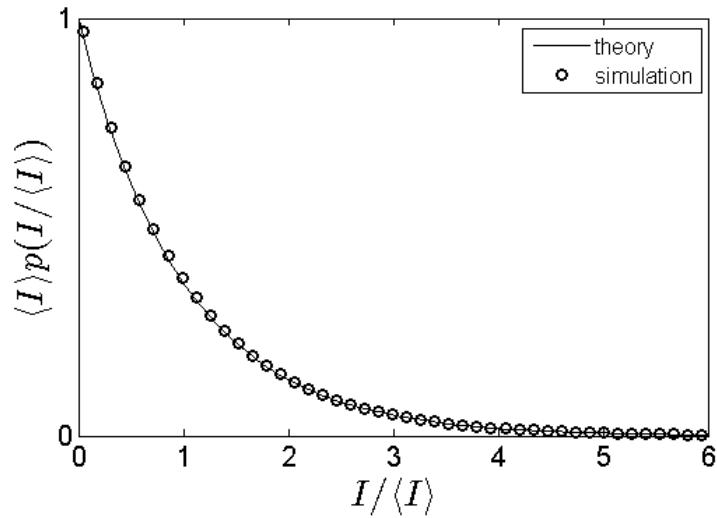

**Figure S1: Speckle intensity distribution.** Theoretical (solid line) and numerical (circles) probability density function of the speckle intensity $I$.

*Speckle correlation function and its Gaussian approximation*

The normalized spatial autocorrelation function of the speckle, which provides a measure of the average speckle grain size, is defined by the diffraction process that generates the speckle itself [3, 15]. For a fully developed speckle, in the general case, this autocorrelation function can be approximated by a Gaussian function whose standard deviation depends on the size of the average speckle grain. In what follows, we treat the case of a speckle pattern generated by diffraction through a circular aperture, as used in the simulations. In this case, the autocorrelation function is the Airy disk (Figure S2):

$$C_I(\Delta \mathbf{r}) = \frac{\langle I(\mathbf{r} + \Delta \mathbf{r}) I(\mathbf{r}) \rangle}{\langle I(\mathbf{r})^2 \rangle} = \left| 2 \frac{J_1(a|\Delta \mathbf{r}|)}{a|\Delta \mathbf{r}|} \right|^2, \quad (S2)$$

where $I(\mathbf{r})$ is the speckle intensity as a function of the position $\mathbf{r}$, $J_1$ is the Bessel function of the first kind and order 1 and $a = \frac{2\pi NA}{\lambda}$, being $\lambda$ the wavelength of light, and $NA$ the numerical aperture under which the speckle is generated. The distance between $|\Delta \mathbf{r}| = 0$ and the first minimum of the Airy function ($|\Delta \mathbf{r}| = 0.61 \frac{\lambda}{NA}$) defines the average speckle diameter at the plane of observation, where $NA$ for a speckle generated at distance $z$ from an area of diameter $D$ is, around the optical axis if $D \gg z$, $NA = n_m \sin\left[\operatorname{atan}\left(\frac{D}{2z}\right)\right]$, where $n_m$ is the refractive index of the medium where the speckle is observed. In our simulations, $D = 1$ mm, $z = 10$ µm, $n_m = 1.33$, and $\lambda = 1064$ nm, giving an average speckle grain size of 490 nm.

Any Airy function is very well approximated by a Gaussian function of standard deviation $\sigma = 0.21 \frac{\lambda}{NA}$ so that [16]

$$C_I(\Delta \mathbf{r}) \approx e^{-\frac{|\Delta \mathbf{r}|^2}{2\sigma^2}}, \tag{S3}$$

or, making explicit the dependence on $x$ and $y$

$$C_I(x, y) \approx e^{-\frac{x^2+y^2}{2\sigma^2}}. \tag{S4}$$

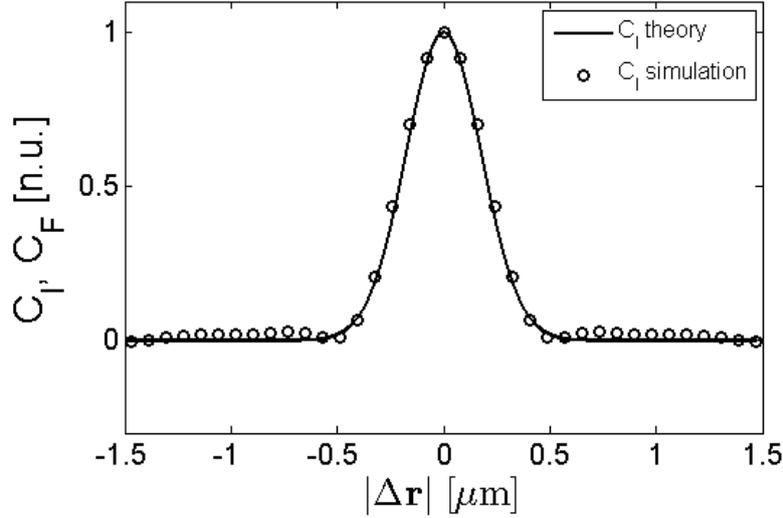

**Figure S2: Speckle intensity autocorrelation function.** Normalized speckle intensity autocorrelation functions as given by the Gaussian model of Equation (S3) (line), and by the Airy function of the numerically generated speckle (circles).

## Supplementary Note S2 – Optical forces in a speckle: probability density function and correlation function

*Force probability density function*

In a speckle field, the optical forces exerted on a Brownian particle are proportional to the gradient of the speckle intensity (see Methods). Since the speckle field is known, we can fully derive numerically the associated random force field and calculate its statistical properties. In particular, from the definition of variance, it can be shown numerically that he following property holds:

$$\langle F^2 \rangle = \langle F \rangle^2 + \text{var}(F) = n\langle F \rangle^2, \quad (S5)$$

where $F$ is the absolute value of the force and $n \approx 1.618$. This property will be useful to derive Equation (S10). In Figure S3, we plot the probability density function of the force for different particle' radii, and, as a guide for the eyes, we fit it to the following empirical function:

$$p(F) = A \left(\frac{nF}{\langle F \rangle}\right)^{\frac{1}{n}} e^{-\frac{nF}{\langle F \rangle}}, \quad (S6)$$

where $A$ is a normalization factor.

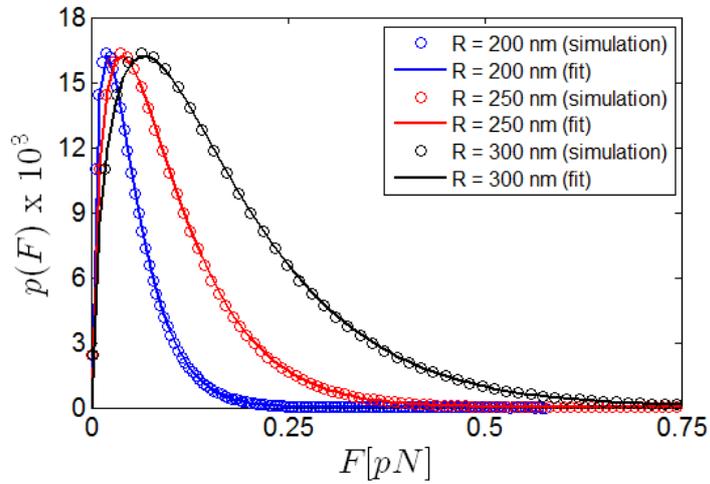

**Figure S3: Probability density function of the optical forces in a speckle field.** Probability density function of the optical forces (circles) acting on particles of different radii moving on the same speckle field, and fitting to Equation (S6) (lines). The average forces increases with the particle radius, $R = 200$ nm (blue circles and line), $R = 250$ nm (red circles and line) and $R = 300$ nm (black circles and line).

*Force correlation function*

The aim is to calculate the force correlation function from the intensity correlation function given in Equation (1).

The correlation of the force can be expressed as

$$\begin{aligned} C_F(\Delta \mathbf{r}) &= \langle \mathbf{F}(\mathbf{r}+\Delta \mathbf{r}) \cdot \mathbf{F}(\mathbf{r}) \rangle \\ &= \langle F_x(\mathbf{r}+\Delta \mathbf{r}) \cdot F_x(\mathbf{r}) \rangle + \langle F_y(\mathbf{r}+\Delta \mathbf{r}) \cdot F_y(\mathbf{r}) \rangle \\ &= C_{F_x}(\Delta \mathbf{r}) + C_{F_y}(\Delta \mathbf{r}) \end{aligned}$$

where $\cdot$ represent the dot-product, $\mathbf{F}$ the force vector, and $F_x$ and $F_y$ the force components. We can now Fourier-transform $C_{F_x}(\Delta \mathbf{r})$ and $C_{F_y}(\Delta \mathbf{r})$ so that

$$\tilde{C}_{F_x}(\mathbf{f}) = \left|\tilde{F}_x(\mathbf{f})\right|^2 = k^2 \left|2\pi i f_x \tilde{I}(\mathbf{f})\right|^2 = -k^2 (2\pi i f_x)^2 \left|\tilde{I}(\mathbf{f})\right|^2,$$

and

$$\tilde{C}_{F_y}(\mathbf{f}) = \left|\tilde{F}_y(\mathbf{f})\right|^2 = k^2 \left|2\pi i f_y \tilde{I}(\mathbf{f})\right|^2 = -k^2 (2\pi i f_y)^2 \left|\tilde{I}(\mathbf{f})\right|^2,$$

Using the Wiener-Khinchin theorem, we can now derive the correlation function of the force from the correlation function of the intensity:

$$C_{F_x}(\Delta \mathbf{r}) = -k^2 \langle I \rangle^2 \frac{\partial^2 C_I(\Delta \mathbf{r})}{\partial x^2} = \frac{k^2}{\sigma^2} \langle I \rangle^2 \left(1 - \frac{x^2}{\sigma^2}\right) e^{-\frac{|\Delta \mathbf{r}|^2}{2\sigma^2}},$$

and

$$C_{F_y}(\Delta \mathbf{r}) = -k^2 \langle I \rangle^2 \frac{\partial^2 C_I(\Delta \mathbf{r})}{\partial y^2} = \frac{k^2}{\sigma^2} \langle I \rangle^2 \left(1 - \frac{y^2}{\sigma^2}\right) e^{-\frac{|\Delta \mathbf{r}|^2}{2\sigma^2}},$$

so that

$$C_F(\Delta \mathbf{r}) = \frac{k^2}{\sigma^2} \langle I \rangle^2 \left(2 - \frac{|\Delta \mathbf{r}|^2}{\sigma^2}\right) e^{-\frac{|\Delta \mathbf{r}|^2}{2\sigma^2}}, \tag{S7}$$

which is rotationally symmetric. From this function we now can extract the force correlation length

$$L = \sqrt{2}\sigma, \qquad (S8)$$

which is the value for which the components of the correlation function have the first zero, so where the force change sign on average.

Moreover, we have

$$\langle F^2 \rangle = C_F(\mathbf{0}) = 2\frac{k^2}{\sigma^2}\langle I \rangle^2, \qquad (S9)$$

and since, from Equation (S5), $\langle F^2 \rangle = n\langle F \rangle^2$, we finally obtain:

$$\langle F \rangle = \sqrt{\frac{2}{n}}\frac{k}{\sigma}\langle I \rangle \approx \frac{k}{\sigma}\langle I \rangle. \qquad (S10)$$

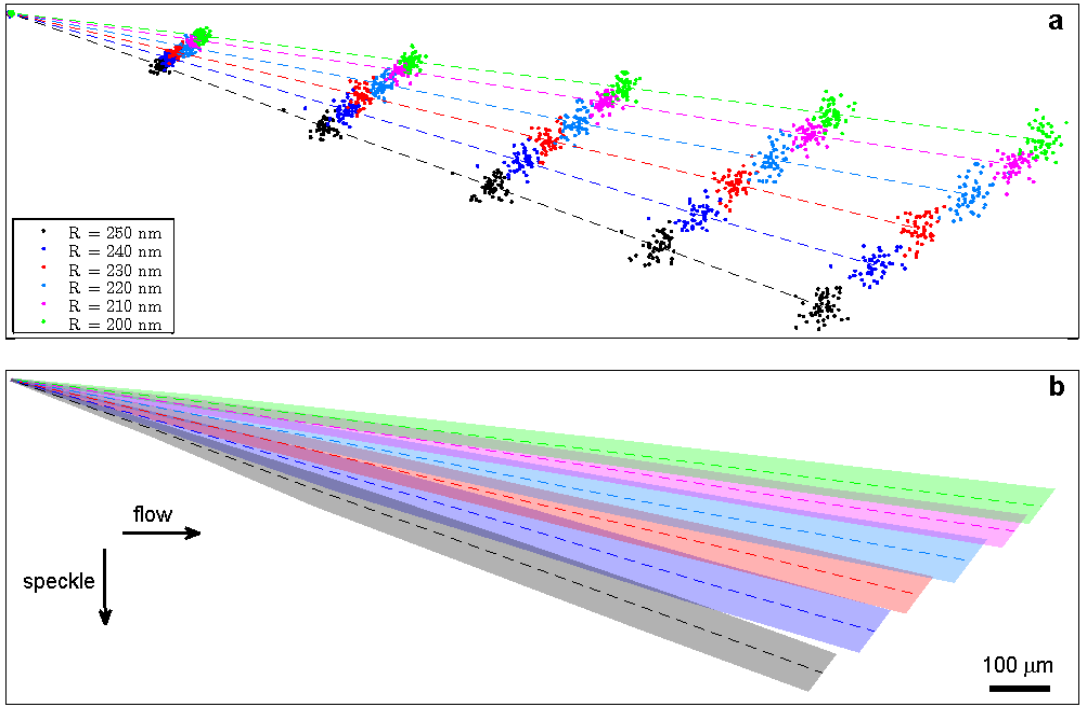

**Figure S4: Fractionation by size.** (**a**) The color-coded symbols represent the positions of particles of various radii ($n_p = 1.59$, radii varying from 250 nm to 200 nm with 10 nm steps) at different times (shown every 10 seconds) as they propagate along a vertically ratcheting speckle in a channel with a flow from the left (flow speed 34 µm/s). The dashed lines represent the motion of the center of mass for the different groups of particles. (**b**) The shaded areas represent one standard deviation of the particle's spread around the mean value. The black scale bar corresponds to 100 µm. The resolution of the fractionation is only limited by the size of the speckle, i.e., the longer the speckle the higher the sensitivity in particle's size.

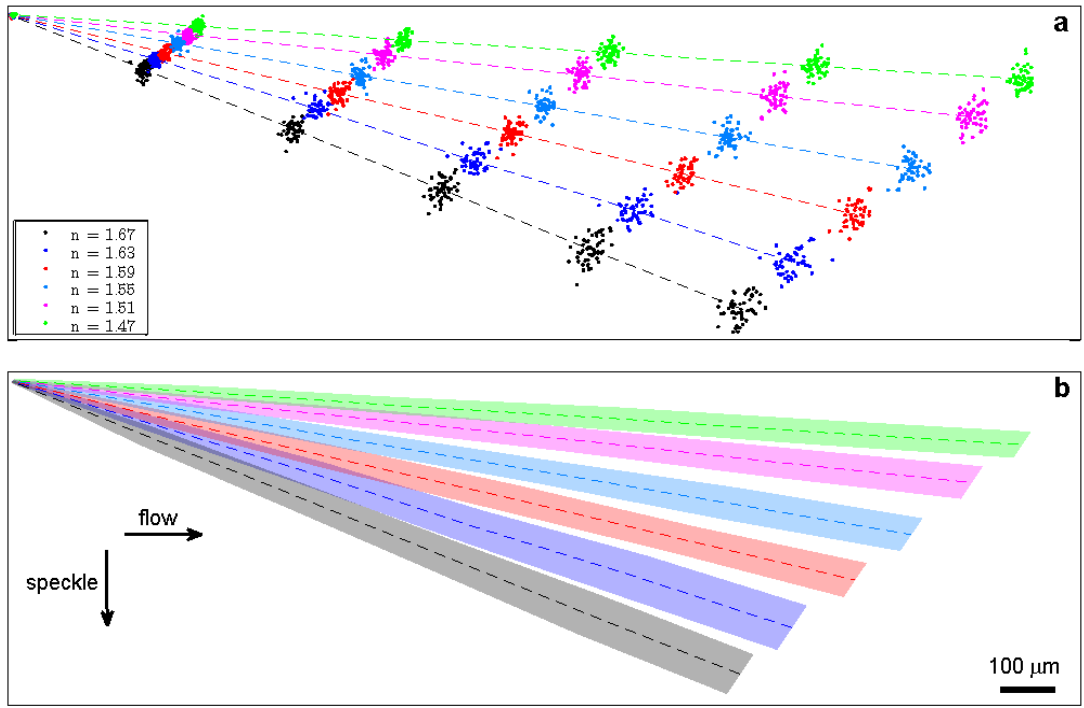

**Figure S5: Fractionation by refractive index.** (**a**) The color-coded symbols represent the positions of particles of various refractive indexes ($R = 225$ nm, $n_p$ varying from 1.67 to 1.47 with 0.04 steps) at different times (shown every 10 seconds) as they propagate along a vertically ratcheting speckle in a channel with a flow from the left

(flow speed 34 μm/s). The dashed lines represent the motion of the center of mass for the different groups of particles. (**b**) The shaded areas represent one standard deviation of the particles' spread around the mean value. The black scale bar corresponds to 100 μm. The resolution of the fractionation is only limited by the size of the speckle, i.e., the longer the speckle the higher the sensitivity in particle's refractive index.